\documentclass[12pt,draftcls,journal,onecolumn]{IEEEtran}

\usepackage{graphicx}
\usepackage{cite}
\usepackage{amsmath}
\usepackage{amssymb}
\usepackage{array}
\usepackage{subfigure}
\usepackage{url}
\usepackage{multirow}
\usepackage{algorithmic}
\usepackage{algorithm}

\usepackage{cuted, flushend}

\usepackage{algorithmic}
\usepackage{algorithm}
\newfloat{program}{thp}{lop}
\floatname{program}{Program}
\usepackage{footnote}

\newcolumntype{C}{>{\centering\arraybackslash} m{3cm} }

\begin{document}

\title{Cost Efficient High Capacity Indoor Wireless Access:
Denser Wi-Fi or Coordinated Pico-cellular?
}
\author{Du Ho Kang, Ki Won Sung, and Jens Zander\\
KTH Royal Institute of Technology, Wireless@KTH, Stockholm, Sweden\\
Email: \{dhkang, sungkw, jenz\}@kth.se}
\maketitle

\begin{abstract}
Rapidly increasing traffic demand has forced indoor operators to deploy more and more Wi-Fi access points (APs). As AP density increases, inter-AP interference rises and may limit the capacity. Alternatively, cellular technologies using centralized interference coordination can provide the same capacity with the fewer number of APs at the price of more expensive equipment and installation cost.  It is still not obvious at what demand level more sophisticated coordination pays off in terms of total system cost. To make this comparison, we assess the required AP density of three candidate systems for a given average demand: a Wi-Fi network, a conventional pico-cellular network with frequency planning, and an advanced system  employing multi-cell joint processing. Numerical results show that dense Wi-Fi is the cheapest solution at a relatively low demand level. However, the AP density grows quickly at a critical demand level regardless of propagation conditions. Beyond this ``Wi-Fi network limit", the conventional pico-cellular network works and is cheaper than the joint processing in obstructed environments, e.g., furnished offices with walls. In line of sight condition such as stadiums, the joint processing becomes the most viable solution. The drawback is that extremely accurate channel state information at transmitters is needed.


\end{abstract}


\section{Introduction}
Wireless data traffic has rapidly increased over the last few years triggered by mobile handset innovation, e.g., tablet PCs or smartphones. The trend of traffic growth is even accelerated by bandwidth-hungry applications, e.g., real-time cloud computing or web-storage service. A significant investment in indoor network deployment is anticipated in upcoming years because most of such traffic increase is expected to come from indoors~\cite{2012smallcell}. An interesting question to operators is what will be the cheapest indoor deployment strategy for such high-capacity provisioning.

Wi-Fi access points~(APs) based on IEEE 802.11 standard have been so far successful due to its significantly lower unit cost than other cellular solutions. However, the contention-based system may lead to a negative impact on overall network capacity because of the fundamental lack of interference coordination. On the other hand, advanced joint processing techniques, often dubbed as networked MIMO, recently emerge as one evolutionary path in the cellular track~\cite{Karakayali2006,Jing2008,Boudreau2009,Gesbert2010}. Theoretically, it can perfectly remove interference, greatly increasing spectral efficiency. Nonetheless, this requires costly dedicated optical fibers for the fast and accurate exchange of channel state information~(CSI) and user data between APs\footnote{Without the loss of generality, we use a term AP for any type of fixed transmitters, e.g., either stand-alone type or remote radio head, at the remainder of paper instead of using base station which is often used in a cellular literature.} which may outweigh equipment costs. Conventional pico-cellular systems with moderate interference coordination, e.g., static resource partitioning, are placed in the middle both from cost and performance perspectives. While tighter interference coordination in the cellular system certainly provides required network capacity with the fewer number of APs, it is still not so straightforward if it will have economic gain over the denser Wi-Fi network in terms of total system cost.
Moreover, the wide variety of in-building structures creates different interference characteristics so that the performance benefit of coordination may diminish in some obstructed environments~\cite{book02rappaport}.
Therefore, quantitative comparison among the candidate solutions is an important research task in various indoor environments.

In order to make the comparison, the both cost and performance of a system should be estimated under the same local environment. In techno-economic literature, we find only the limited number of cost comparisons between a Wi-Fi and a conventional pico-cellular system that are based on empirical cost figures from a specific market~\cite{Jan10iofc,Markendahl2011,phdthesis07johansson,KA07pimrc}.
These studies mostly relied on simplified performance estimation which disregarded the distinct interference coordination mechanisms, e.g., CSMA/CA randomness in the Wi-Fi. For instance, in~\cite{Jan10iofc,Markendahl2011}, an overall network performance was obtained by scaling average per-cell throughput by the number of placed APs. In~\cite{phdthesis07johansson,KA07pimrc}, it was estimated more precisely by reflecting the statistics of user distribution and propagation conditions.
Still, a link-level data rate was approximated as the function of received average signal to average interference plus noise ratio~(SINR).
%
In contrast, significant technology-oriented studies evaluated performance of either W-Fi or cellular systems with more refined coordination models~\cite{Bianchi2000,Nguyen2007,Alfano2011,Simic2012,Caire2010,Ramprashad2009,Marsch2010,Bosisio2008,Chandrasekhar2009,Andrews2010,Son2011}. For instance, authors in~\cite{Bianchi2000,Nguyen2007,Alfano2011,Simic2012} estimated a performance of a Wi-Fi network by using different mathematical tools. The others in~\cite{Caire2010,Ramprashad2009,Marsch2010} studied the theoretical capacity of the cellular network by the different means of multi-antenna techniques or proposed more practical interference management algorithms~\cite{Bosisio2008,Chandrasekhar2009,Andrews2010,Son2011}. %
However, they have thus far overlooked the importance of analyzing the economic benefit of cellular systems than the Wi-Fi system for a given data rate requirement.
This led the performance of each system to be assessed in parallel in different environments with unsynchronized working assumptions and performance metrics.

The objective of our study is providing an insight into the potential way towards the future lowest-cost indoor deployment by explicitly comparing both Wi-Fi and cellular systems which have distinct interference coordination mechanisms and cost structures.
%
While exact cost figures for equipments and backhaul are market and vendor specific, estimating the required AP density to meet an average traffic demand is our way to compare total network cost of systems which is approximately linear to the AP density~\cite{JS97VTC}.
For comparison purpose, the AP density of three systems is quantitatively assessed for a given average traffic demand: a Wi-Fi network, a conventional pico-cellular network with frequency planning, and an advanced cellular network with an emerging joint processing technique.
The main contribution of this work is three-fold:
\begin{enumerate}
\item First, the required AP density of the three aforementioned systems is estimated as a function of increasing average user demand under outage constraint, particularly taking into account technical differences in the interference coordination. The impact of indoor propagation conditions, i.e., indoor wall loss and pathloss exponent, are further examined in order to see how the economic gain of one system deviates accordingly.
\item Secondly, we develop a framework for evaluating the optimistic performance of a Wi-Fi network which is applicable to obstructed environments but still retaining the CSMA/CA mechanism. In the framework, an ideal MAC and PHY layer are explicitly modeled based on random sequential packing process~\cite{Busson2009,A.Busson2009,Viet2012} which maximally exploits spatial reuse without redundant collisions in order to estimate the minimally required Wi-Fi AP density.
\item Thirdly, we provide a solution map of future indoor deployment which intuitively guides operators to choose the lowest-cost option. It is characterized by traffic demand and environmental openness with identified practical bottlenecks.
\end{enumerate}
The rest of this paper is organized as follows.
Section II provides an overall system model including network dimensioning and traffic assumptions. Then, Wi-Fi and cellular network models are presented in Section III and IV, respectively. Section V describes a simulation methodology to estimate an individual system performance. Numerical results illustrate implications on the economic indoor system in Section VI. Finally, we draw conclusions in Section VII.

\section{System Model}

\subsection{Service Area Layout}
\begin{figure}
{
    \centering
    \includegraphics[trim=0cm 3cm 0cm 2cm,clip=true,     width=0.8\columnwidth]{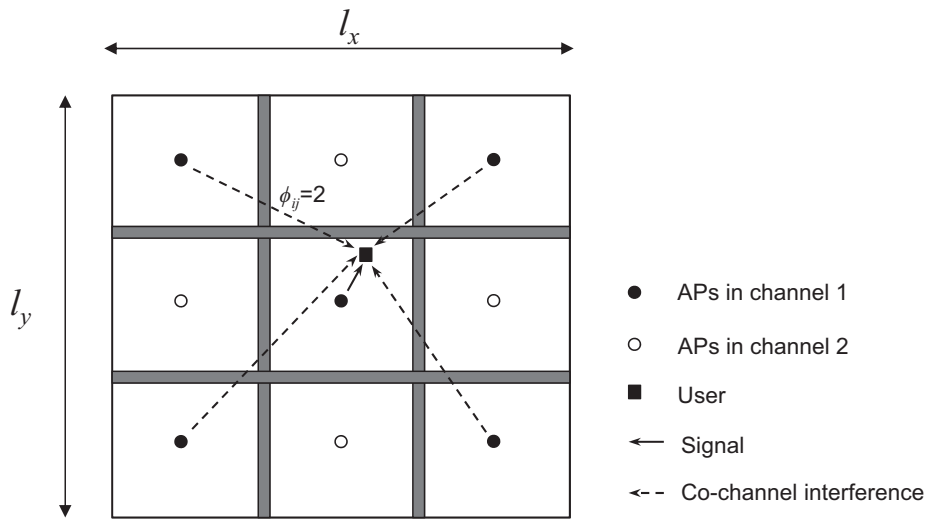}
    \caption{Service area layout with the example of AP deployment in nine rooms.}
    \label{fig:building}
}
\end{figure}


As depicted in Fig.~\ref{fig:building}, we consider a finite square service area $\Omega$ with the size of $l_x$ by $l_y$ in meters. We assume that there are walls which are equally placed and create the constant signal strength loss $L_w$~(dB). This represents the homogenous indoor materials for simplicity.
The numbers of walls are $w_x$ and $w_y$, which are placed vertically and horizontally, respectively, in the following locations:
\[{x_w} = \frac{{{l_x}a}}{{{w_x} + 1}}~~{\rm{and}}~~{y_w} = \frac{{{l_y}b}}{{{w_y} + 1}},\]
where $a=1, ..., w_x$ and $b=1, ..., w_y$.
%

\subsection{Traffic Assumptions}
We focus on downlink traffic
which is the main consideration of network deployment.
Mobile users equipped with single antenna are assumed to be uniformly and independently distributed with the average density of $E[\lambda_u]$ (users/$\rm{km^2}$).
Typically, the traffic load pattern varies drastically on an hourly basis~\cite{Forum2011}. Thus, a busy hour, which refers to the time during a day with the most traffic, is a commonly used concept in telecommunications for network dimensioning~\cite{KA07pimrc,Johansson2007vtc}.
We assume that the $\omega$ fraction of the day are busy hours when the traffic is intensified.
By assuming that all users are equally served during the busy hour in an average sense, a network can support a feasible average user throughput denoted by ${\mu}$ (Mbps/user) during the busy hour.
%
In addition, average monthly traffic volume per user in GB/month/user has been used as the conventional metric of measuring network usage in telecommunication business since it provides an intuitive understanding of customer behavior. For instance, the mobile broadband usage today is 1-2 (GB/month/user) in most developed countries~\cite{2012ericsson_traffic}.
Following this convention, our work adopts this metric instead of classical system spectral efficiency in bps/Hz which is common in most of technical analysis.
Then, average user demand $\mathcal{D}$ (GB/month/user) can be defined by
\[
\mathcal{D}:= \frac{c_0}{\omega}{\mu}~(\rm{GB/month/user}),
\]
where $c_0$ represents the scaling constant to transform Mbps/user unit to GB/month/user. It is $\frac{1}{{1024}} \times \frac{{1}}{{8}} \times 3600 \times 30$ by assuming $30$ days a month. Although average user demand per month strongly depends on user statistics in local places, i.e., $\omega$ and $E[\lambda_u]$, the comparison of deployment density is still valid since it is mainly affected by system performance and physical propagation conditions.



%

\subsection{Propagation Model}
We assume that both  an AP and a user have single antenna for the analysis simplicity. Then, each radio link is affected by the path loss attenuation and Rayleigh fading. For the path loss between AP~$i$ and user~$j$, we adopt a general Indoor Multiwall and Floor model which considers all walls intersecting the direct ray between an AP and a user~\cite{Lott2001,book02rappaport}. By assuming single floor, the path loss between AP~$i$ and user~$j$ can be dependent on the internal wall and distance as given in
\[
L^{(\rm{dB})}_{ij}=L_0+10\alpha log_{10}(d_{ij})+\phi_{ij}L_w~(\rm{dB}),
\label{eq:path}
\]
where $L_0$, $\alpha$, $d_{ij}$, and $\phi_{ij}$ represent the constant loss, a pathloss exponent, the distance in meter between AP~$i$ and user~$j$, and the number of walls across a AP~$i$ and a user~$j$, respectively.
Also, $\alpha$ depends on the size or the surroundings of the rooms or operating frequency. In general, the bigger size of rooms with hard obstacles at higher frequency creates the higher $\alpha$~\cite{book02rappaport,Tolstrup2008}.
%

Additionally, we assume an independent and identical small-scale Rayleigh fading channel component $z_{ij}\sim\mathcal{CN}(0,\sigma^2_z)$. By assuming rich scattering, all channels are perfectly uncorrelated. This can be justified in a fact that APs or users are sufficiently separated and many scattering objects in indoors~\cite{Foschini1998a}, e.g., furniture or uneven wall surfaces.
Then, let us define the complex channel response between AP~$i$ and user~$j$ as
\[
h_{ij}=\sqrt{L_{ij}}z_{ij}.
\]
Note that $L_{ij}$ represents the linearly scaled path gain, i.e.,  $L_{ij}=10^{\frac{-L^{(\text{dB})}_{ij}}{10}}$. Then, a complex channel coefficient matrix is denoted by $\rm{\bf{H}}=\{h_{ij}\}$ for all active links in a given time.
Then, the channel power gain $g_{ij}$ can be computed from squaring the amplitude of $h_{ij}$, i.e., $g_{ij}=|h_{ij}|^2$.
We also consider a block fading model, where $\rm{\bf{H}}$ remains quasi-static within a fading block, but varies between contiguous fading blocks.

\subsection{Performance Measure}
The overall network capacity is evaluated with respect to average area throughput density E[$\lambda_s$] (Mbps/$\rm{km}^2$) subject to the outage probability constraint $\nu = \rm{Pr}(\rm{SINR}_j < \gamma_t)<\beta$.
$\gamma_t$ represents the minimum SINR for feasible transmission. Note that $\lambda_s=\sum R_j$~(Mbps/$\rm{km}^2$) represents aggregate data rates of all served users per unit area which is governed by an involved coordination mechanism and the channel realization.
For average user density $E[\lambda_u]$~(users/$\rm{km}^2$), we assume that a deployed network can support $\mu$ (Mbps/user) during the busy hour on average:
\[
E[\lambda_s]=\mu E[\lambda_u]~(\rm{Mbps/}\rm{km}^2).
\]
%

\subsection{Network Dimensioning}


The APs should be dimensioned to support $\mu$ during the busy hour for a given average user density $E[\lambda_u]$ under the constraint of  $\nu < \beta$. We assume that the APs are placed regularly in order to maintain the equal geographical area per BS. That is, $n_x$ by $n_y$ APs are placed as shown in Fig.~\ref{fig:building}. The location planning could be further optimized considering the building geometry and its indoor structure. Since our objective is comparing systems employing the distinct interference coordination mechanisms, we leave the location optimality issue out of the scope. Instead, we refer interested readers to our previous work on this aspect~\cite{Kang2011self}. We simply assume the locations $(x_{\rm{ap}},y_{\rm{ap}})$ of deployed APs are given as
\[(x_{\rm{ap}},y_{\rm{ap}}) = \left( {\frac{{{l_x}}}{n_x}\left( {\frac{1}{2} + (a - 1)} \right),\,\frac{{{l_y}}}{n_y}\left( {\frac{1}{2} + (b - 1)} \right)} \right)\]
where $a=1, ..., n_x$ and $b=1, ..., n_y$.
\section{Wi-Fi Network Model}
The performance evaluation of a Wi-Fi network itself is a challenging task since the mathematical modeling of the complex behavior of multiple APs is difficult. An accurate packet-level simulation is also very time-consuming particularly in a dense network. Thus, we propose a Wi-Fi network model where practical impairments in MAC and PHY layer are idealized, but the key features of CSMA/CA operation are captured. This model can be used to estimate the optimistic performance of a multi-channel Wi-Fi network in obstructed indoor environments to yield the minimal number of APs for a given user demand.

\subsection{802.11 Channel Model}

Depending on the IEEE 802.11 standard variants, a number of orthogonal frequency channels are defined. The 802.11b/g standard provides a set of 14 frequency channels among which only 3 non-overlapping channels are possible, whereas 802.11a offers around 8 non-overlapping channels depending on the spectrum availability in 5 GHz~\cite{ieee802.11}.
%
Thus, the frequency reuse number $K^{\rm{wifi}}$ for Wi-Fi is different counting on the operating frequency and a specific standard.
Then, we assume that each AP~$i$ access one of non-overlapping channels whose bandwidth $w^{\rm{wifi}}_i$ is given for the total system bandwidth $W$ by
\[
w^{\rm{wifi}}_i=\frac{W}{K^{\rm{wifi}}}~\rm{(MHz)}.
\]

\subsection{MAC and PHY Layer Model}

In principle, the MAC operation is based on CSMA/CA mechanism regardless of the operating frequency and standard variants. We also assume that all APs transmit data at the same fixed power $P_t$~(mW).
Since we suppose that a network is dimensioned to meet average user throughput during a busy hour, persistent downlink traffic is presumed with no or negligible upstream traffic. 
Let us define a set of APs operating in a frequency channel $k$ as $\mathcal{A}^k$. 
As for the association, each user attaches to the AP offering the highest $L_{ij}$.
Note that each user may instantaneously have worse signal strength from the associated AP than nearby APs due to fading component $z_{ij}$. However, this assumption reflects the practical association criteria which is based on average signal strength.

At a given data transmission time,
only a subset of $\mathcal{A}^k$ can be concurrently active due to CSMA/CA operation.
Let $\mathcal{A}_x^k$ be the set of all co-channel APs which are in the contention domain of AP~$x$ operating in a channel $k$. It is mathematically expressed as follows:
\[
  \mathcal{A}_x^k:=\{i\in\mathcal{A}^k, i \neq x |g_{ix}P_t>CS^{li}_{thr}\},
\]
where $CS^{li}_{thr}$~(mW) represents the linearly scaled carrier sensing threshold of $CS_{thr}$~(dBm), i.e., $CS^{li}_{thr}=10^{\frac{CS_{thr}}{10}}$. The collision probability (outage in a cellular context) and the spatial reuse gain are affected by $CS_{thr}$ since it defines the size of the contention domain.
Intuitively, the higher $CS_{thr}$ makes APs aggressively reuse the space domain, i.e., more simultaneous transmission for higher capacity but with more collision. Typically, the IEEE 802.11 standard suggests $CS_{thr}$ for equal transmission opportunity in unlicensed band~\cite{ieee802.11}. For this, we use $CS_{thr}$=-85~dBm as defined in \cite{ieee802.11} which is named as \textit{Baseline} Wi-Fi. Nevertheless, there may be a strong motivation for operators to adjust $CS_{thr}$ for striking their own balances between capacity and collision probability. Thus, we also examine lower $CS_{thr}$ for the comparison purpose which is referred to as \textit{Aggressive} Wi-Fi.

In a practical Wi-Fi network, collisions among APs in a contention domain can occur because of the mismatch in the timing of carrier sensing and actual transmission. However, we assume that all APs in $\mathcal{A}_x^k$ always listen when AP~$x$ transmits data.
Although the perfect carrier sensing removes the impairment due to temporal mismatch, collisions due to spatial mismatch between carrier-sensing APs and receiving users, a so-called hidden node problem, can still occur.
However, the impact of the hidden node problem becomes negligible  with denser deployment because the distance between a user and its serving AP is much closer than the carrier-sensing range~\cite{Nguyen2007}. Likewise, the conservative transmission due to an exposed node problem also diminishes with densification.

Let us refer to the set of all active APs elected by the MAC protocol in a frequency channel $k$ at a given time as $\Phi^k\subset \mathcal{A}^k$. We here omit the time index $t$ unless any ambiguity is introduced. Since $\Phi^k$ plays an important role in the performance of a Wi-Fi network, a precise and yet simple model describing this set is necessary while capturing the important features of CSMA/CA.
%
Mat\'{e}rn hard core point process has been recently paid an attention to estimate data rates of a generic CSMA/CA network. This is theoretically founded on Poisson point process~(PPP) with an additional thinning process~\cite{Cardieri2010}. Several studies used this to model the MAC operation for estimating data rates of a dense Wi-Fi network~\cite{Nguyen2007,Alfano2011}.
However, the classical homogeneous PPP assumption suffers from several drawbacks in practical indoor situations, as argued in~\cite{Haenggi2005,Ganti2006,Cardieri2010}.
First of all, the effect of obstructed indoor environment, which creates spatially varying interference, is hard to be analyzed even if it is one of crucial factors to affect the performance of indoor deployment.
Secondly, it probabilistically permits some APs out of the contention domains to be silent. This leads to the reduced spatial reuse gain, i.e., fewer number of simultaneous transmission.
Thirdly, the spatial average of AP locations in infinite service areas is estimated by assuming homogeneous interference. Practical indoor Wi-Fi networks have finite service areas such that users near the boundary experience lower interference level than those at the center.
Lastly, it is only applicable to single channel operation which overestimates the collision probability than multi-channel Wi-Fi deployment in practice.

\begin{figure}
{
    \centering
    \includegraphics[trim=0cm 3cm 0cm 3cm,clip=true, width=1\columnwidth]{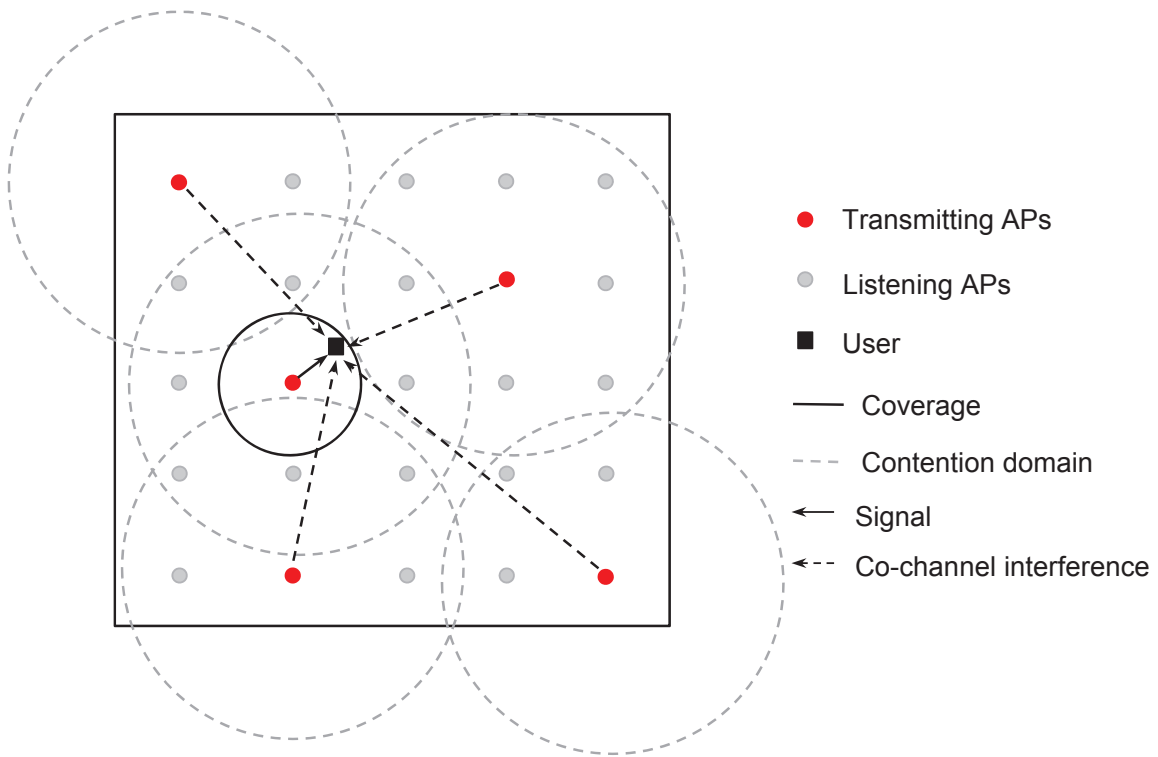}
    \caption{An example of realization of active Wi-Fi APs when single channel is available without any walls.}
    \label{fig:wifi}
}
\end{figure}

These motivate us to consider a Simple Sequential Inhibition (SSI) process which is applicable to more realistic indoor environments by empirical simulation~\cite{Viet2012}.
As a family of random sequential packing process, the SSI is known to be more appropriate to model CSMA/CA networks and was originally developed for a finite continuous spatial domain~\cite{Busson2009,A.Busson2009}. Let us define a SSI process $\Psi$ in a constructive manner on a finite and discrete set $\mathcal{A}^k$ rather than arbitrary points in the service area. $\Psi(n)$ is composed of a sequence of $n$ random variables denoted by $X_1$, ..., $X_n$ which are independently and uniformly distributed in $\mathcal{A}^k$. Initially, $X_1$ is added to $\Psi(1)$. Then, $X_i$ is systematically added to $\Psi(i)$ if and only if it does not belong to the contention domain of any APs in $\Psi(n-1)$, i.e., $X_i \notin \cup_{X_j \in \Psi(i-1)} \mathcal{A}^k_{X_j}$. The process stops whenever entire APs in $\mathcal{A}^k$ are either active APs or in the contention domain of any active APs.
A sample of active APs $\Phi_k$ can be built by such SSI process as exemplified in Fig.~\ref{fig:wifi}. This process always lets APs transmit unless it is in the range of the contention domain of other active APs, contrary to Mat\'{e}rn point process.
Note that a recent work in~\cite{Simic2012} also proposed a simulation-based model to approximate a data rate in a multi-channel dense Wi-Fi network. Although the data rate is modeled as a function of AP density, it still assumed the pessimistic interference where any active APs always experience interference form all APs in $\mathcal{A}^k$.

For a given realization $\Phi^k$, each active AP $x\in \Phi_k$ randomly selects user $j$ to transmit data. Then, the data rate can be ideally achieved as
\begin{multline}
\label{eq:wlan}
    R_j^{\rm{wifi}}=\min \Bigg\{w^{\rm{wifi}}_i \log_2\Bigg(1+\frac{g_{ij}P_t}{\sum_{x\in \Phi^k\backslash i}g_{xj}P_t+\sigma^2/K^{\rm{wifi}}}\Bigg),R^{\rm{wifi}}_{max}\Bigg\}~\rm{(Mbps)},
\end{multline}
where $R^{\rm{wifi}}_{max}=w^{\rm{wifi}}_i \eta^{\rm{wifi}}$ and $\sigma^2$~(mW) is average noise power for whole system bandwidth~$W$.
A practical Wi-Fi system uses a discrete rate and the rate adaptation is also imperfect due to the lack of explicit channel information feedback from users unlike the cellular systems~\cite{ieee802.11}. In this regard, the rate model implicitly assumes the perfect rate adaptation without any quantization. Thus, our approach again gives us the optimistic performance of the dense Wi-Fi network.

\section{Cellular Network Model}
For comparative analysis with the Wi-Fi network, we consider two representative indoor cellular systems with distinct coordination techniques: a conventional pico-cellular system with static frequency planning and an advanced system employing multi-cell zero-forcing~(ZF) beamforming.

\subsection{Conventional Frequency Planning}
The frequency planning is the one of static coordination techniques which were used in a conventional cellular network to mitigate multi-cell interference. In the conventional planning, whole system bandwidth $W$ is partitioned into $K$ non-overlapping equal-width bands in a centralized planning process.
%
Then, each AP~$i$ transmits at full power $P_t$ using one out of $K$ disjoint bands whose amount is given as:
\[
  w^{\rm{sta}}_i=\frac{W}{K}~(\rm{MHz}).
\]
Recall that $K^{\rm{wifi}}$ in the Wi-Fi network is fixed irrespective of propagation conditions due to the restrictions in standards whereas $K$ in the conventional cellular coordination can be adaptively determined.
%
For a given $K$, a user~$j$ is randomly selected at a given time and achieves data rate of $R_j^{\rm{sta}}$ as
\begin{multline}
\label{eq:sta}
    R_j^{\rm{sta}}=
    \min \Bigg\{w^{\rm{sta}}_i \log_2\Bigg(1+\frac{g_{ij}P_t}{\sum_{x\in \mathcal{A}^k\backslash i}g_{xj}P_t+\sigma^2/K}\Bigg),R_{max}^{\rm{sta}}\Bigg\}~\rm{(Mbps)},
\end{multline}
where $R_{max}^{\rm{sta}}=w^{\rm{sta}}_i \eta^{\rm{sta}}$.
There are also other variants in the static coordination approach such as fractional frequency reuse~\cite{Stolyar2008,Boudreau2009}. However, we confine our study to the static frequency planning as a reference since our objective is to compare representative cellular systems with the Wi-Fi system.


\subsection{Multi-cell Zero Forcing Beamforming}
While co-channel interference still exists in the conventional pico-cellular network, it can be ideally removed by the means of advanced multi-cell joint processing techniques. Several precoding strategies in the joint processing have been proposed and studied~\cite{Gesbert2010,Karakayali2006}. In general, dirty paper coding~(DPC) is known as the precoding strategy to yield the information theoretical upper bound~\cite{Costa1983,Yu2007}.
Alternatively, suboptimal linear precoding based DPC, commonly referred to as ZF-DPC, deals with the part of interference with a linear precoding strategy whereas remaining interference still relies on nonlinear DPC coding~\cite{Caire2003}. However, it is generally very complex to implement such non-linear precoding techniques in a practical system. Instead, zero forcing~(ZF) beamforming which is entirely a linear processing is considered both in academia and industries as one way of practical implementations of the joint processing~\cite{Karakayali2006,Yoo2006b,Somekh2009,2010TR_CoMP}.
The ZF technique becomes near-optimal in high SNR regime which is at the focus of our interest, i.e., a dense indoor network. Thus, we will employ the linear ZF as the representative joint-processing technique.

Although the ZF has a great potential to improve the performance of the indoor network, it may be sensitive to imperfect channel estimation or feedback delay~\cite{Caire2010,Ramprashad2009,Marsch2010}.
Therefore, we consider two cases for the ZF: ideal ZF with perfect CSI at transmitters~(CSIT) and one with CSIT errors.

\subsubsection{Ideal ZF}
Let us refer to the vector of signals transmitted by $N$ single antenna APs as $\rm{\bf{x}}\in \mathbb{C}^{[N\times 1]}$. Then, the vector of received signals $\rm{\bf{y}}\in \mathbb{C}^{[M\times 1]}$ in $M$ single antenna users can be described as
\[
\rm{\bf{y=Hx+n}},
\]
where $\rm{\bf{H}}\in \mathbb{C}^{[M\times N]}$ and $\rm{\bf{n}}\in \mathbb{C}^{[M\times \bf{1}]}$ is referred to as an additive white noise vector with covariance $E[\rm{\bf{n}}\rm{\bf{n}}^\dag]=\sigma^2 \rm{\bf{I}}$.

Suppose that $\rm{\bf{H}}$ is perfectly known at the network side without any feedback delay and estimation/quantization errors.  $\rm{\bf{x}}$ can be now reconstructed with linear beamforming matrix $\rm{\bf{W}}\in\mathbb{C}^{[N\times M]}$ as:
\[
\rm{\bf{x=Wu=H^{\dag}(HH^{\dag})^{-1}u}},
\]
where the $j$-th element in the vector $\rm{\bf{u}}\in \mathbb{C}^{[M\times 1]}$ is the information-bearing data symbol intended for the $j$-th user.
We also assume that the elements of $\rm{\bf{u}}$ are
independent zero-mean complex Gaussian random variables
with variance $E[u_j u_j^{\dag}]=p_j$~(mW).
Then,
\[
\rm{\bf{y=Hx+n=HH^{\dag}(HH^{\dag})^{-1}u+n=u+n}}.
\]
User~$j$ receives $y_j=u_j+n_j$ so that it can achieve data rate
\begin{equation}
\label{eq:zf}
    R_j^{\rm{zf}}=\min \Bigg\{W \log_2\Bigg(1+\frac{p_j}{\sigma^2}\Bigg), R^{\rm{zf}}_{max}\Bigg\}~\rm{(Mbps)},
\end{equation}
where $R^{\rm{zf}}_{max}=W\eta^{\rm{zf}}$ represents the maximum link data rate supported by the network.
For the fair comparison with other systems, we disregard the multiuser diversity gain by randomly selecting single active user per AP in a given time.
Thus, all channel matrices considered in the sequel are $N$-by-$N$ by setting $N=M$.
Because we assume that data symbols heading to different users follow independent and zero-mean Gaussian distribution, per-antenna power constraint (PAPC) can be expressed as linear constraints, i.e. $E[|x_i|^2]=\sum_j |w_{ij}|^2p_j\leq P_{t}$. Moreover, aggregate rates $\sum_j {{R^{\rm{zf}}_j}}$ are concave in the symbols power vector $p_j$.
Then, power allocation for the sum rate maximization becomes a convex programming problem which is efficiently solvable by standard optimization techniques~\cite{Boyd2004}.

\subsubsection{$\delta$-erroneous ZF}
%

When ZF is used in a practical cellular network, the interference may not be perfectly removed since some elements of $\rm{\bf{H}}$ can be outdated due to the channel dynamics and feedback delay. This leads to inaccurate symbol power allocation and the beamforming matrix $\rm{\bf{W}}$ selection.
%
We propose a simple method to model this by introducing the probability $\delta$ that CSIT is outdated for a given link. We assume that only fading component $z_{ij}$ varies in consecutive fading blocks with consistent $L_{ij}$. Then, at a given feedback delay $\tau$ between channel estimation and data transmission at time $t$, let us define $\delta$ as

\[\delta : = {\rm{Pr}}\left( {{\rm{|}}z_{ij}^{t} - z_{ij}^{t-\tau }{\rm{|}} > {\rm{0}}} \right).\]
If the CSIT is outdated, we also assume that there is correlation of $\rho$ between $z_{ij}^{t}$ and its delayed version $z_{ij}^{t-\tau}$. This is modeled as the first order of autoregressive process~\cite{Baddour2005}:
\[
z_{ij}^{t}=\rho z_{ij}^{t-\tau}+\sqrt {1 - {\rho ^2}} q_{ij}^{t}
\]
where $q_{ij}^{t}\sim{\cal C}{\cal N}(0,\sigma_z^2)$ is independently and identically distributed for arbitrary links.
We here consider CSIT inaccuracy only due to feedback delay by latency of backhaul and CSI report.
Including other practical barriers for using ZF, e.g., estimation errors and pilot channel/CSIT report overhead in the air, may bring out additional performance degradation. Nevertheless, analyzing one of dominant barriers still allows us to investigate the feasibility of ZF in practice.
%

%
%

\section{Simulation Methodology}

Recall that our objective is the quantitative comparison of the required AP densities of three wireless systems to satisfy a certain traffic demand. Thus, estimating $E[\lambda_s]$ according to different number of deployed APs is essential.
In this section, we provide a simulation methodology to estimate $E[\lambda_s]$ subject to the outage constraint $\nu <\beta$ by facilitating a snap-shot based Monte-Carlo simulation.

\subsection{Numerical Optimization for Resource Allocation}
For a certain $K^{\rm{wifi}}$, the frequency assignment in a Wi-Fi network needs to be optimized. Traditionally, the multi-cell frequency assignment was formulated as a graph-coloring problem. However, it is generally known as a NP-hard problem due to its combinatorial property~\cite{Gesbert2007}. Thus, when the large number of APs are deployed, a search for the optimal solution is prohibitive. Instead, we employ a heuristic algorithm where each AP is randomly and sequentially chosen and is allocated the frequency channel generating the minimum aggregate interference.
In the frequency planned cellular network, both the frequency reuse number $K$ and its assignment should be optimized. In the classical Wyner model where hexagonal cell patterns are infinitely repeated, $K$ can be determined by the closed-form of expressions~\cite{Zander2001}. However, it cannot be applicable to our indoor layout which is composed of finite service areas with indoor walls. Instead, we exhaustively search the lowest $K^*$ subject to $\nu<\beta$ by evaluating network performance for all possible $K$. The network performance for each possible $K$ is evaluated by applying the same frequency assignment algorithm used for the Wi-Fi network. While the frequency assignment to the Wi-Fi and conventional cellular system is performed based on average propagation loss, the symbol power allocation in ZF is optimized in every snapshot of channel realization by using a convex optimization package to exploit the dynamic coordination~\cite{cvx2011}.

\subsection{Simulation Procedure}
When APs are deployed, $K^{\rm{wifi}}$ and $K$ orthogonal frequency channels are assigned in the Wi-Fi network and the conventional cellular network, respectively. Then, a snapshot based Monte-Carlo method is applied to estimate the average performance of three systems.
In each snapshot, a simulation procedure is described as follows:
\begin{enumerate}
\item A set of users are uniformly and independently dropped and each user associates with the AP providing the strongest average signal strength. Each AP randomly selects one user to be served. The channel gain matrix $\rm{\bf{H}}$ is generated for the chosen users.
\item For the case of a Wi-Fi network, one realization of $\Phi_k$ is generated by the SSI process to create the active set of APs. The optimal symbol power in ZF is also computed and allocated for selected users.
\item Then, we evaluate Eq.~(\ref{eq:wlan}), (\ref{eq:sta}), and (\ref{eq:zf}) for all served users and the sample values of aggregate data rates $\lambda_s$ in each system and count the number of active users in outage.
\end{enumerate}
We estimate $E[\lambda_s]$ and $\nu$ by averaging the independent sample values. Afterward, we enumerate $E[\lambda_s]$ by increasing the number of APs and apply the same simulation procedure to different $L_w$ and $\alpha$.


\section{Numerical Results}

%

%
In this section, we provide numerical results obtained by simulation with the parameters summarized in Table~\ref{tab:sys_parameter}. We illustrate the required AP density which meets the average user demand under the outage probability constraint. This is further examined in two different propagation conditions: the line of sight~(LoS) conditions without any walls and higher pathloss exponent with walls.

\begin{savenotes}
\begin{table}
\caption{Simulation Parameters} \centering
\label{tab:sys_parameter}
\begin{tabular}{|m{7cm}|C|}
\hline
\bfseries Parameter Name & \bfseries Values \\
\hline
\hline Service area size & $l_x=l_y$=100~m \\
\hline Traffic intensity portion during a busy hour $\omega$ & 0.2 \\
\hline Average number of users $E[\lambda_u]$ & $\rm{10^5}$ users/$\rm{km^2}$\\
\hline Constant pathloss $L_0$ & 37~dB \\
\hline The maximum link spectral efficiency in the Wi-Fi system\footnote{We assume $\eta^{wifi}$ achievable in popular 802.11a/b/g systems which support up to 54~Mbps for 20~MHz bandwidth.} $\eta^{\rm{wifi}}$ & 2.7~bps/Hz\\
\hline The maximum link spectral efficiency in the cellular system\footnote{It is assumed that $\eta^{\rm{sta}}$ and $\eta^{\rm{zf}}$ are equivalent to the link spectral efficiency required in IMT-Advanced systems in single antenna configuration~\cite{2008itureq}} $\eta^{\rm{sta}}$, $\eta^{\rm{zf}}$ & 3.75~bps/Hz \\
\hline Boltzmann constant $k$ & $\rm{1.38\times10^{-23}}$~J/K\\
\hline Temperature $T$ & 300~K\\
\hline Fading variance $\sigma_z^2$ & 1 \\
\hline Correlation factor in feedback delay$\rho$ & 0.9 \\
\hline Average transmission power $P_t$ & 100~mW \\
\hline The minimum required SINR $\gamma_t$ & 3~dB \\
\hline The minimum outage probability constraint $\beta$ & 0.05 \\
\hline The number of available non-overlapping channel for the Wi-Fi\footnote{We implicitly assume three non-overlapping channel as 802.11b/g in 2.4 GHz.} $K^{\rm{wifi}}$ & 3 \\
\hline
\end{tabular}
\end{table}
\end{savenotes}

\subsection{Deployment Options in Open Environment}


In this subsection, we compare the deployment density of the Wi-Fi network and the other two cellular solutions when $\alpha$=2 and $L_w=0$, which is more likely to occur with network densification.
\subsubsection{Wi-Fi Network Limit}
\begin{figure}
{
    \centering
    \includegraphics[width=0.8\columnwidth]{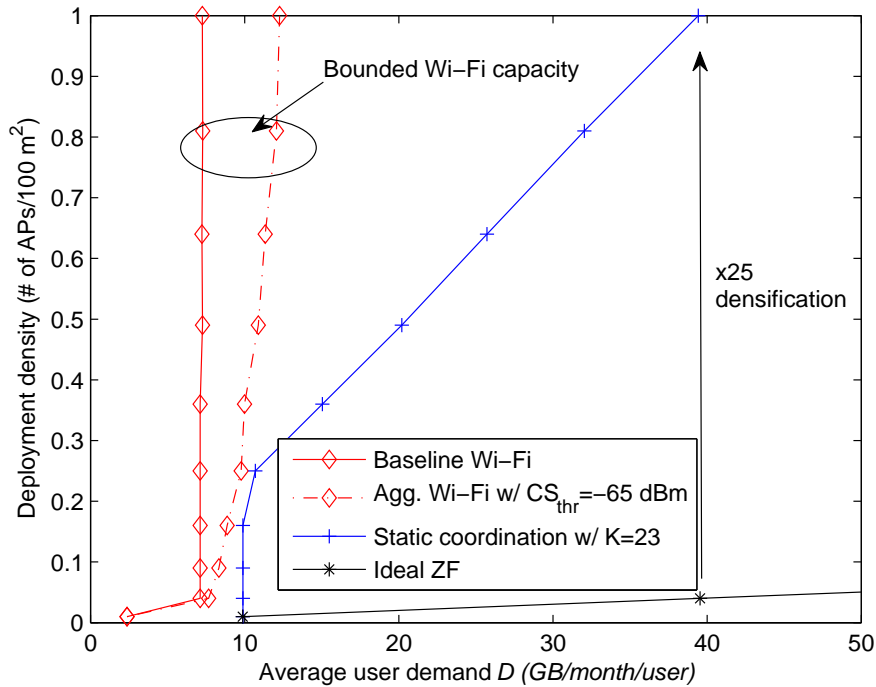}
    \caption{The comparison of deployment density in an open environment ($\alpha$=2, $L_w$=0 dB).}
    \label{fig:dem_dens_2}
}
\end{figure}

\begin{figure}
{
    \centering
    \includegraphics[width=0.8\columnwidth]{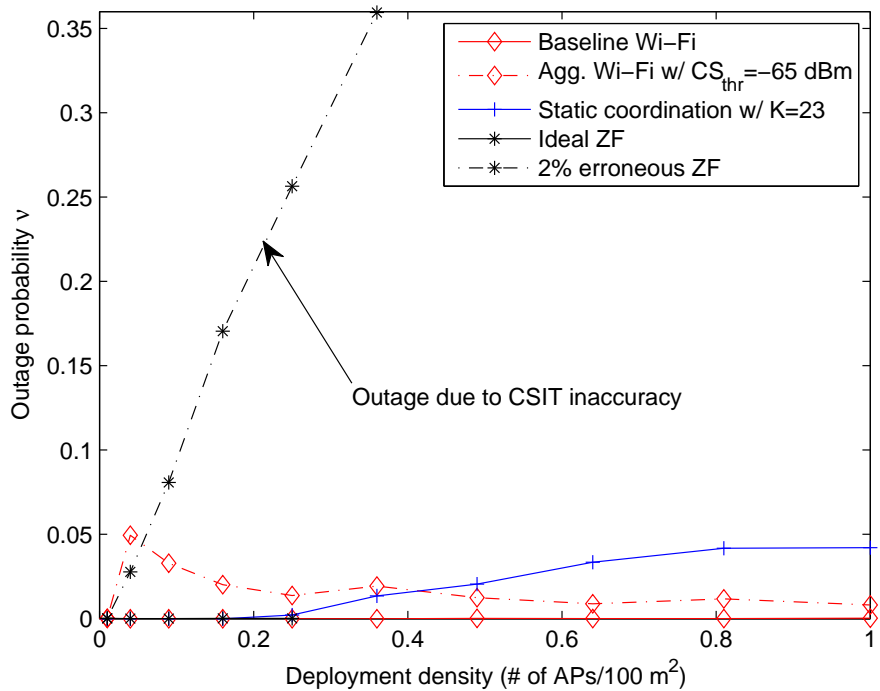}
    \caption{Outage probability according to densification ($\alpha$=2, $L_w$=0 dB).}
    \label{fig:out_dens_2}
}
\end{figure}

We can identify from Fig.~\ref{fig:dem_dens_2} that the baseline Wi-Fi densification has almost no capacity expansion. This can be explained as follows.
Since all co-channel APs in $\Omega$ are in the contention domain in our parameter setting, only one AP in each channel can be active in a given time regardless of AP density. Thus, there can be three active APs at maximum which is equal to $K^{\rm{wifi}}$. In addition, every served user can reach the highest rate $R_{max}^{\rm{wifi}}$ regardless of its location in $\Omega$ because high received signal strength can be achieved due to very low propagation loss. Consequently, the densification does not result in network capacity expansion, i.e., no gain from spatial reuse and shorter service range.
%
In the aggressive Wi-Fi with $CS_{thr}$=-65~dBm, extra capacity is gained with more collisions (less than $\beta$).
It is noticeable in Fig.~\ref{fig:out_dens_2} that the densification decreases the collision since it lets APs stay closer to their users while interference is maintained by $CS_{thr}$. Still, the deployment density rapidly grows again when average user demand exceeds a certain level, i.e., roughly 7~GB/month/user. Beyond this level, additional Wi-Fi densification marginally increase the network capacity so that alternative cellular solutions are essential.

\subsubsection{Practical Challenge in Networked MIMO}
When compared with the conventional cellular network, the multi-cell ZF beamforming ideally provides the same capacity with much less AP density which was corroborated in many existing studies~\cite{Karakayali2006,Yoo2006b,Somekh2009}. For instance, it requires 25 times less densification than conventional cellular network when user demand is 40~GB/month/user. In total cost perspective, the joint processing can be economic unless deployment cost per AP is 25 times more expensive than the conventional cellular system.
However, only $2\%$ errors in $\rm{\bf{H}}$ results in very large outage and it is magnified with densification for the given $\delta$ as shown in Fig.~\ref{fig:out_dens_2}. Accordingly, extremely high accuracy on CSIT is required to maximally exploit the cost benefit of the joint processing.

\subsubsection{Spectrum Bottleneck in Static Coordination}

\begin{figure}
{
    \centering
    \includegraphics[width=0.8\columnwidth]{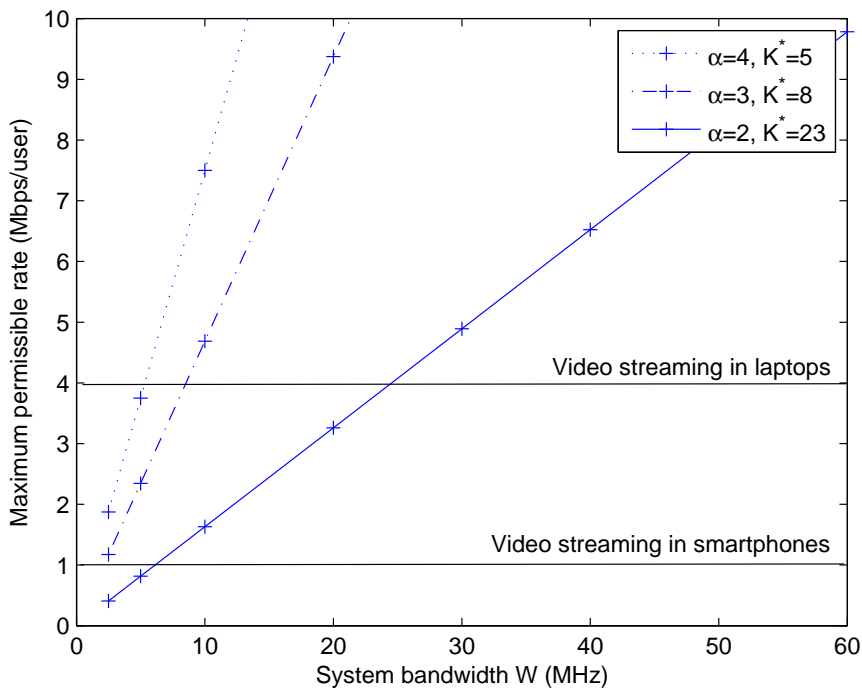}
    \caption{The maximum rate supported by an AP according to system bandwidth $W$ (1~AP/100~$\rm{m}^2$, $L_{w}=0$~dB).}
    \label{fig:percell}
}
\end{figure}

If the multi-cell ZF may not be feasible due to imperfect CSIT, the conventional cellular network could be a only remaining solution for high-capacity indoor service.
However, the maximum data rate may be limited to support future high data rate services although densification of the conventional cellular system can increase the average network capacity.
For instance, video streaming services, as one of main drivers for traffic growth in near future~\cite{cisco2010}, requires the various ranges of consistent minimum data rate according to the type of devices and resolution as provided in Table~\ref{table:video}. Fig.~\ref{fig:percell} illustrates the maximum permissible data rate with the optimal $K^*$ depending on available total system bandwidth~$W$ and $\alpha$.
Even without any waste of spectrum usage for guard band protection or other protocol overhead, we can find in the LoS condition that more than 5~MHz is required to serve a single smartphone user whereas typical laptops at least need 25~MHz.
Therefore, spectrum will be one practical bottleneck in the conventional static coordination approach in order to
provision such high-bandwidth real-time service in open environments.
\begin{savenotes}
\begin{table}
\caption{Examples of typical bit rate for video streaming ~\cite{2009videoLTE}}
\centering  
\begin{tabular}{c|c|c|c} 
\hline                  
Device type & Screen size & Typical& Typical H.264/\\
            & & screen resolution & MPEG4 Bit rate\footnote{These rates include typical IP overhead, but do not include audio (which may add a few 10's of Kbps depending on audio encoding)} \\
\hline\hline                        
Smartphones & 3-3.5$^{\prime\prime}$ & 480$\times$320 & 500 K to 1~Mbps \\
\hline
PMP & 4-7$^{\prime\prime}$ & 800$\times$480 & 800 K to 1.5~Mbps \\
\hline
Tablet PCs & 7-9$^{\prime\prime}$ & 800$\times$480,  & 1 to 2~Mbps \\
 &                      & 1024$\times$600 & \\
\hline
Laptops & 12-17$^{\prime\prime}$ & 1280$\times$800, & 3 to 4~Mbps \\
        &                        & 1920$\times$1200 & \\
\hline 
\end{tabular}
\label{table:video} 
\end{table}
\end{savenotes}

\subsection{Deployment Options in Obstructed Environment}

\begin{figure}
{
    \centering
    \includegraphics[width=0.8\columnwidth]{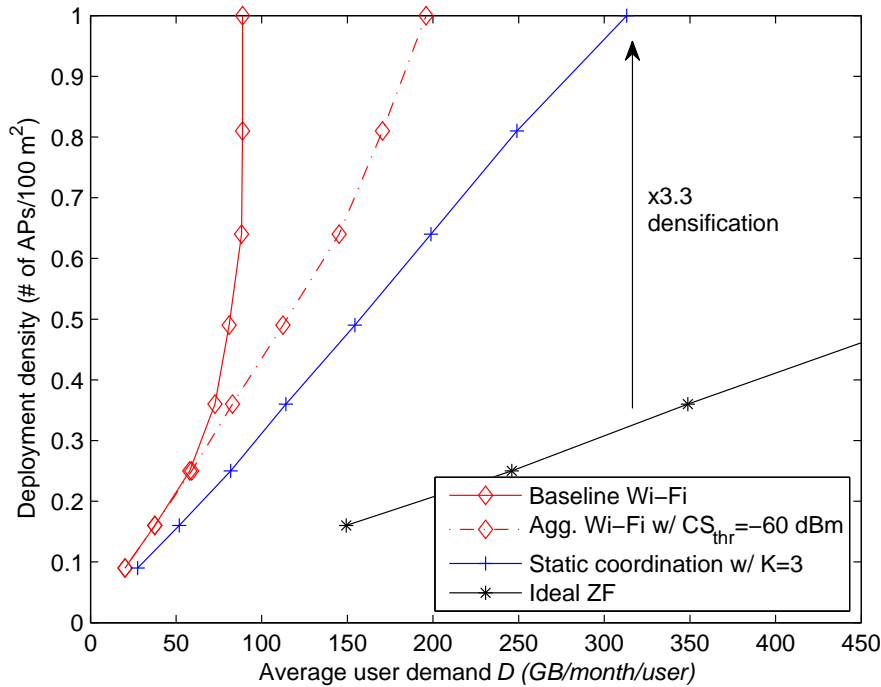}
    \caption{Deployment density comparison in a closed environment ($\alpha$=4, $L_w$=10 dB with $w_x$=$w_y$=4).}
    \label{fig:dem_dens_4}
}
\end{figure}
\begin{figure}
{
    \centering
    \includegraphics[width=0.8\columnwidth]{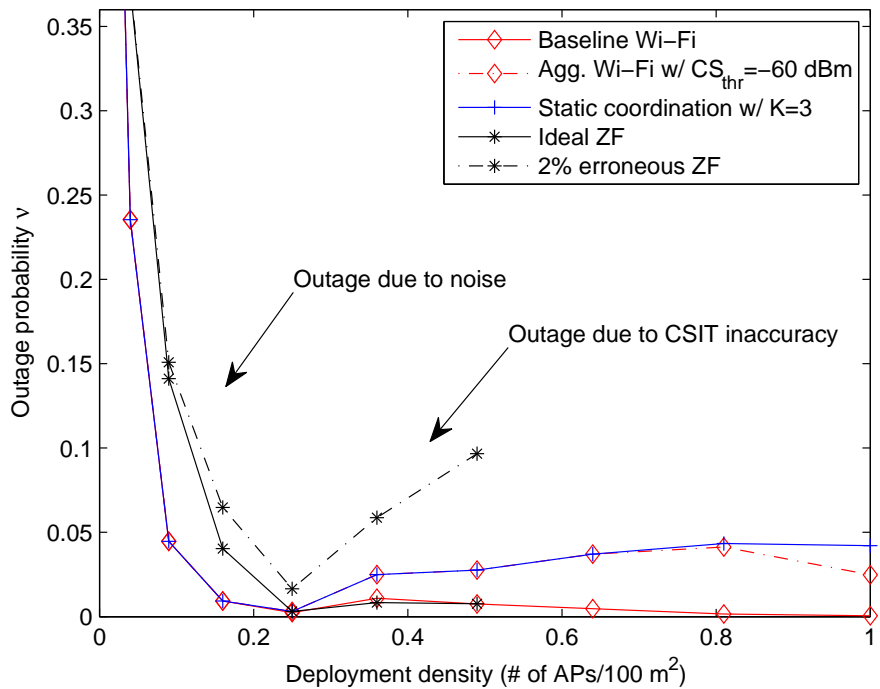}
    \caption{Outage probability according to deployment density ($\alpha$=4, $L_w$=10 dB).}
    \label{fig:out_dens_4}
}
\end{figure}

While the LoS condition is likely to occur in wide open areas, e.g., shopping malls or outdoor hotspots on streets, there are also indoor environments with many walls and higher $\alpha$, e.g., furnished offices~\cite{book02rappaport}. Thus, the impact of different propagation conditions on the deployment options needs to be studied. To make a contrast from the harsh interference condition in the previous subsection, we consider a propagation condition where 25 rooms are introduced with $\alpha=4$. Fig.~\ref{fig:dem_dens_4} provides the quantified deployment density in this environment.
The baseline Wi-Fi densification can support more user demand thanks to better spatial reuse, i.e., reduced contention domain, than the open environment.
Accordingly, the range of user demand where a Wi-Fi can stay as the lowest cost solution is extended.
Still, the required AP density rapidly grows after a certain demand level, i.e., when the number of APs exceed the number of rooms for the baseline Wi-Fi. In the end, the Wi-Fi network capacity is bounded due to the fundamental limitation in CSMA/CA.
Similarly to the open environment, aggressive Wi-Fi can give more capacity with increased $\nu<\beta$ as illustrated in Fig.~\ref{fig:out_dens_4}. Nevertheless, the AP density begins to rapidly increase in very high demand.


Beyond the Wi-Fi network limit, the ideal ZF can satisfy the user demand with fewer AP density than conventional frequency planning.
As shown in Fig.~\ref{fig:out_dens_4}, it is noticeable that the outage of ideal ZF is higher with low AP density than other system solutions. This is because reduced signal strength due to the energy spent to penetrate walls for interference cancellation may outweigh the gain of removing interference. The required ratio of densification for the frequency planned network over the multi-cell ZF significantly drops from 25 to 3.3 when we compare Fig.~\ref{fig:dem_dens_2} and Fig.~\ref{fig:dem_dens_4}. In the presence of the given cost for the advanced equipment and optical fibers installation, this indicates that the joint processing becomes a much less attractive solution in obstructed environments. The environmental impact on the performance benefit of the joint processing has been neglected in most of existing work despite the importance of considering wide variety of indoor environments. More discussions on this issue can be found in our previous work~\cite{Kang2012a}.

\subsection{Implications for Affordable Future Indoor Deployment}
 \begin{figure}
{
    \centering
    \includegraphics[trim=0cm 3.5cm 0cm 4.5cm, clip=true, width=0.85\columnwidth]{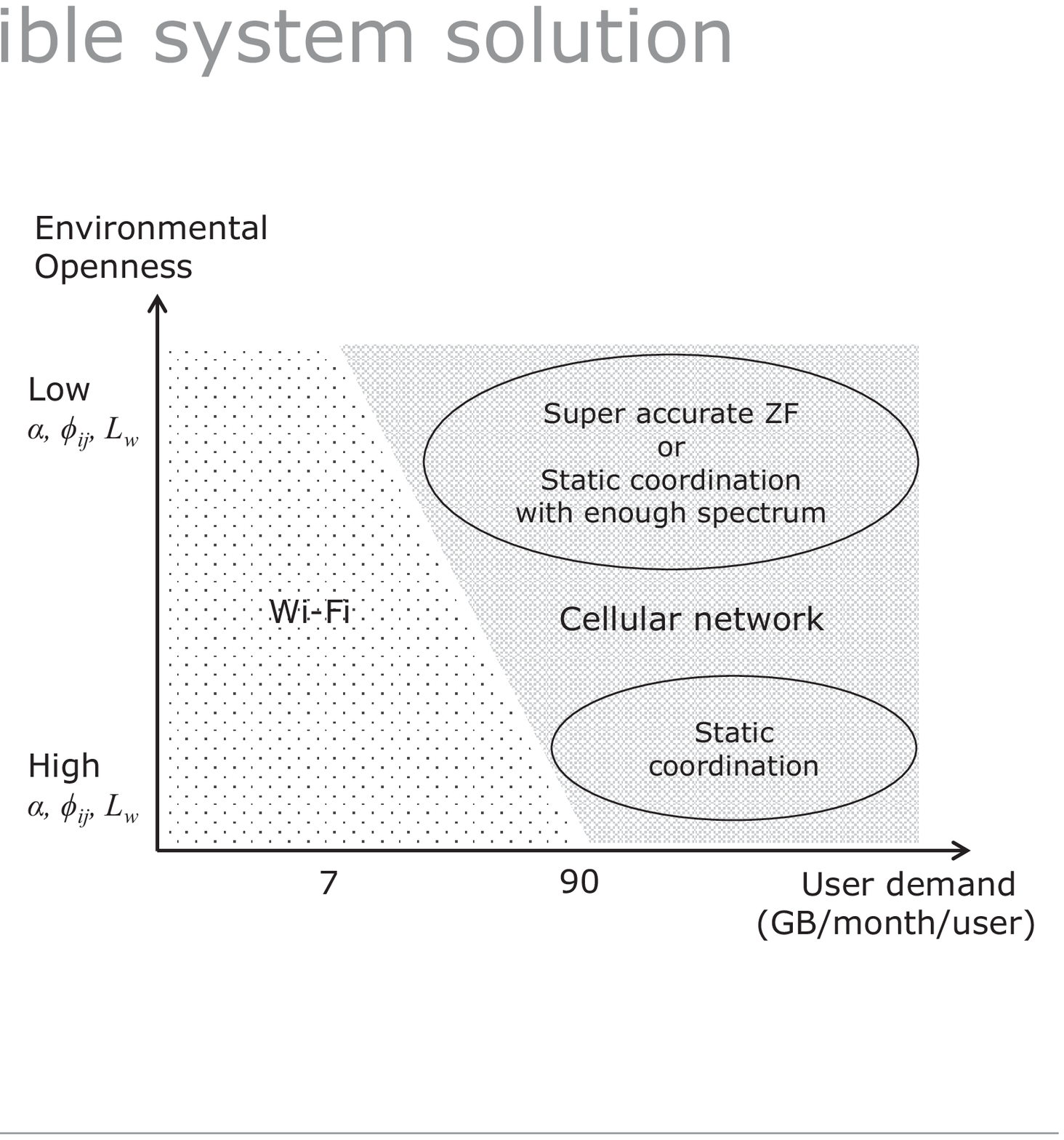}
    \caption{A solution map for an economic indoor wireless system.}
    \label{fig:concl}
}
\end{figure}

Based on our findings, we can bring out a map for the cheapest wireless system solution in Fig.~\ref{fig:concl}. It is guided by two crucial factors, i.e., environmental openness characterized by $\alpha$, $\phi_{ij}$, $L_w$ and average user demand in GB/month/user. It well separates the region for three representative wireless systems with identified practical bottlenecks. In low user demand, an existing Wi-Fi network can be the economic solution due to its low AP equipment cost and almost free backhaul cost. However, as the demand increases, denser AP deployment cannot support it due to limited spatial reuse gain from the CSMA/CA mechanism. In the high demand, the conventional cellular solution with static coordination can be the cheapest option particularly in closed environments, i.e., high $\alpha$ and rich walls such as offices. This is because the joint processing may not have sufficient performance gain to compensate the expenses of pricey fiber installation. However, in open environments such as concert halls or department stores where the LoS condition is likely to occur, the joint processing can support the high demand with cheaper system cost than the static coordination since it can have substantial gain from interference cancellation. This can be enabled only when extremely high CSIT accuracy can be achieved. Otherwise, sufficient amount of spectrum needs to be secured so that the conventional cellular network can guarantee the quality of service (QoS) of high-speed services.

\section{Conclusion}
We studied the economic benefit of interference coordination for high-capacity indoor deployment in terms of total network cost. We quantitatively compared the required AP density of three representative systems at a given average traffic demand: a Wi-Fi network, a conventional pico-cellular network with static frequency planning, and an advanced cellular network with multi-cell zero-forcing beamforming.
We found that the cost benefit of interference coordination strongly depends on average user demand level and propagation conditions.
Wi-Fi densification can be the lowest cost deployment option as of now when expected demand is relatively low in obstructed environments with high pathloss exponent and walls. Nonetheless, after a certain demand level, Wi-Fi deployment cost rapidly grows due to the fundamental limitation of time sharing mechanism in CSMA/CA. Beyond such a Wi-Fi network limit, the conventional pico-cellular network can be the cheapest solution in the obstructed environments because the performance gain of the joint processing considerably drops in the presence of wall loss. The joint processing is identified as the only viable solution in open environments. However, extremely accurate CSIT needs to be assured for achieving the performance gain in practice.
%
In this work, we assumed a fully-loaded network during busy hour which leads to pessimistic interference situations. A more realistic interference model reflecting the dynamic fluctuation of traffic load should be further considered to estimate network performance.

\bibliographystyle{IEEEtran}
\bibliography{H:/BACK_UP_moved_to_H_drive/KTH_Research/BibTex/BibTeX-dhkang-Abrv,H:/BACK_UP_moved_to_H_drive/KTH_Research/BibTex/BibTeX-dhkang-Items2}

\end{document}